\documentclass[preprint,superscriptaddress,nofootinbib,amsmath,amssymb]{revtex4-2}

\usepackage{ulem}

\usepackage{graphicx}
\usepackage{subcaption}
\usepackage[colorlinks,linkcolor=magenta,anchorcolor=cyan,citecolor=blue]{hyperref}
\usepackage{physics}
\usepackage{tikz}
\usepackage[justification=raggedright]{caption}

\begin{document}

\title{Search for the non-linearities of gravitational wave background in NANOGrav 15-year data set}

\author{Jun-Qian Jiang}
\email{jiangjunqian21@mails.ucas.ac.cn}
\email{jiangjq2000@gmail.com}
\affiliation{School of Fundamental Physics and Mathematical
    Sciences, Hangzhou Institute for Advanced Study, UCAS, Hangzhou
    310024, China}
\affiliation{School of Physical Sciences, University of
Chinese Academy of Sciences, Beijing 100049, China}

\author{Yun-Song Piao}
\email{yspiao@ucas.ac.cn}
\affiliation{School of Fundamental Physics and Mathematical
    Sciences, Hangzhou Institute for Advanced Study, UCAS, Hangzhou
    310024, China}
\affiliation{School of Physical Sciences, University of
Chinese Academy of Sciences, Beijing 100049, China}
\affiliation{International Center for Theoretical Physics
    Asia-Pacific, Beijing/Hangzhou, China}
\affiliation{Institute of Theoretical Physics, Chinese
    Academy of Sciences, P.O. Box 2735, Beijing 100190, China}

\begin{abstract}
The recently reported signal of common red noise between pulsars
by several pulsar timing array collaborations has been thought as
evidence of the stochastic gravitational wave background (SGWB)
due to the Helling-Downs correlation. In this Letter, we
search for the non-Gaussianity of SGWB through its non-linear
effect on the overlap reduction function in NANOGrav 15-year data
set. In particular, we focus on a folded component to SGWB whose
amplitude is quantified with a single parameter $\alpha$ in the
unpolarized case.
% { \color{red}
The resulting Bayes factor of $1.68 \pm 0.01$ ($1.78 \pm 0.01$ in the case of signals from SMBHBs) indicates that there is no evidence of such a non-Gaussianity of SGWB in the NANOGrav 15-year data yet.
% }
% The results reveal that such a non-Gaussianity
% of SGWB is slightly favored with a Bayes factor of $1.68 \pm 0.01$ ($1.78 \pm 0.01$ in the case of signals from SMBHBs),
If it is detected in future PTA experiments, it will impact our understanding on
the origin of the detected SGWB.

\end{abstract}

\maketitle
\newpage

\section{Introduction} Several pulsar timing array (PTA)
collaborations have recently reported strong evidence for the
signal of the stochastic gravitational wave background (SGWB)
\cite{NANOGrav:2023gor,Antoniadis:2023rey,Reardon:2023gzh,Xu:2023wog}.
They have found the Helling-Downs correlation
\cite{Hellings:1983fr} between pulsars, which is the ``smoking
gun" for the SGWB. These results suggest the SGWB observed might
be primordial GW
\cite{Vagnozzi:2023lwo,Vagnozzi:2020gtf,Benetti:2021uea,
Jiang:2023gfe,Piao:2004tq,Piao:2004jg,Liu:2011ns,Liu:2012ww,
Zhu:2023lbf,Cai:2016ldn,Cai:2015yza,Cai:2020qpu,Ye:2023tpz}, or
produced by the merger of binary black holes and other
possibilities, e.g.\cite{Ellis:2023dgf,Huang:2023chx,Bi:2023tib,
Choudhury:2023kam,Oikonomou:2023bli,Datta:2023xpr,
Wang:2023len,Ghosh:2023aum,Li:2023bxy,Liu:2023ymk,Niu:2023bsr,DiBari:2023upq,
Wang:2023sij,Du:2023qvj,Ye:2023xyr,Balaji:2023ehk,Zhang:2023nrs,Ellis:2023oxs,
Li:2020cjj,An:2023jxf,Xiao:2023dbb,HosseiniMansoori:2023mqh,King:2023ayw,He:2023ado,Frosina:2023nxu,
Maji:2023fhv,Kawasaki:2023rfx,Kawai:2023nqs,Lozanov:2023rcd,Gangopadhyay:2023qjr,
Chen:2023zkb,Chen:2023uiz,Liu:2023hpw,Choudhury:2023fwk,Choudhury:2023fjs,Huang:2023mwy,
Chen:2024fir,Domenech:2024cjn,Choudhury:2024one,Basilakos:2023xof,Basilakos:2023jvp,Ahmadvand:2023lpp,Inomata:2023zup,Maiti:2024nhv}.

It is well known that the detected SGWB can come from
astrophysical processes or early Universe. But its non-Gaussianity
is usually ignored.
\footnote{
Recent discussions (e.g. \cite{Franciolini:2023pbf}) on ``non-Gaussianity" and PTA-detected SGWB focus on a scalar-induced GW scenario where ``non-Gaussianity" refers to the non-Gaussianity of scalar perturbation, whereas what we focus on here is the non-Gaussianity of SGWB itself.
}
% { \color{red}
This is because the usual non-Gaussianity (higher order correlators) of
inflationary or other cosmological SGWB (see e.g.
\cite{Bartolo:2018qqn,LISACosmologyWorkingGroup:2022jok} for
recent reviews) may not be directly measured due to the phase
decoherence that occurs during the very long propagation of gravitational waves from their emission to the detection~\cite{Bartolo:2018evs,Bartolo:2018rku,Margalit:2020sxp,Kehagias:2024plp}, while only stationary (i.e., folded shape) signals survive due to phase cancellation.
% }
As for the astrophysical processes, the cumulative effect of many
independent sources will produce a Gaussian signal according to
the central limit theorem. However, if GW events are not frequent
enough to overlap in time, i.e. the central limit theorem is not
exactly applicable, the non-Gaussianity of SGWB would be
non-negligible. It might be this case for SGWB from the merger of
sparse binary supermassive BHs \cite{NANOGrav:2023bts} (or binary
supermassive PBHs,
e.g.\cite{Huang:2023chx,Huang:2023klk,Huang:2023mwy}), or with the
louder individual events \cite{NANOGrav:2023pdq}. 
Therefore,
in order to identify the origin of such a SGWB, it is essential to
check the non-Gaussian level on the time scale.

Several methods have been proposed to detect the non-Gaussanity of
the SGWB through PTA experiments
\cite{Tsuneto:2018tif,Powell:2019kid,Tasinato:2022xyq}. Some of
them are focusing on the higher correlator between pulsars, which
is hard to detect at the current level, and the computational cost
is expensive. According to Ref.\cite{Tasinato:2022xyq},
the
non-Gaussanity on the time scale might affect the overlap
reduction function (ORF) and deviate from the Hellings-Downs (HD) curve~\cite{Hellings:1983fr}, which has been detected, through its
non-linear effect. In this Letter, we will search for the
corresponding non-Gaussinity of SGWB in the recent NANOGrav
15-year data set, in particular, we will focus on a folded
non-Gaussian component to SGWB.

%whose amplitude is quantified with a single parameter $\alpha$ in
%the unpolarized case.

%the contributions from the bispectrum vanished, thus we will focus
%on the effect of the trispectrum \footnote{The trispectrum we
%consider here is unpolarized
% (see e.g.\cite{Chen:2021wdo,Chen:2023uiz,NANOGrav:2023ygs} for search
% for the polarization modes)
%and described by a single parameter
%$\alpha$.}

% The methodology and the data we employed are described in
% \autoref{sec:methodology} and we present the results and discuss
% them in \autoref{sec:result}.

\section{Methodology}

\subsection{Non-Gaussianity-induced deviations from HD curves}

In the TT gauge the GW modes are the tensor
perturbations $h_{ij}(t,\overrightarrow{x})$ satisfying
$h^i_i=\partial_i h^{ij}=0$.
%The gravitational wave can be
%expressed as the fluctuation of the spacetime metric in the TT
%gauge:
%\begin{equation}
%    \dd s^2 = - \dd t^2 + (\delta_{ij} + h_{ij}) \dd x^i \dd x^j .
%\end{equation}
For the photon emitted at $t_\text{em}$ and traveling from a
pulsar $a$ to the earth, the relation between the distance and the
time interval is
\begin{equation}
    \dd x^2_a = \frac{\dd t^2}{1 + h_{ij} \hat{x}^i_a \hat{x}^j_a} .
\end{equation}
As a result, the photon will arrive at
\begin{equation}
    t_\text{obs} = t_\text{em} + d_a + \int_{t_\text{em}}^{t_\text{em}+d_a} \dd t' \left[ 1 - \frac{1}{\sqrt{1+E_a}} \right] \left( t', (t_\text{em} + d_a - t')\hat{\mathbf{n}}_a \right)
\end{equation}
where $\hat{\mathbf{n}}_a$ is the unit vector from the Earth
toward the pulsar $a$, and $d_a$ is the distance between them.
Here, we have denoted $E_a = h_{ij} \hat{x}^i_a \hat{x}^j_a$ for
simplicity.

Similarly, for the next photon emitted at $t'_\text{em} =
t_\text{em} + T_a$ after the period $T_a$, it will be observed at
\begin{equation}
    t'_\text{obs} = t_\text{em} + T_a + d_a + \int_{t_\text{em}}^{t_\text{em}+d_a} \dd t' \left[ 1 - \frac{1}{\sqrt{1+E_a}} \right] \left( t' + T_a, (t_\text{em} + d_a - t')\hat{\mathbf{n}}_a \right)
\end{equation}
Consider a monochromatic GW propagating along the direction
$\hat{\mathbf{n}}$, the extra contribution to the period $\Delta
T_a = t'_\text{obs} - t_\text{obs} - T_a$ can be convert to the
relative time delay
\begin{align} \label{eq:za}
    z_a = \frac{\Delta T_a}{T_a} &= - \frac{1}{1 + \hat{\mathbf{n}}_a \cdot \hat{\mathbf{n}}} \left[ \frac{1}{\sqrt{1 + E_a(t, \mathbf{x}=0)}} - \frac{1}{\sqrt{1 + E_a(t-\tau_a, \mathbf{x}_a)}} \right] \\
    &= \frac{1}{2(1 + \hat{\mathbf{n}}_a \cdot \hat{\mathbf{n}})} \left[ (E_a(t, \mathbf{x}=0) - E_a(t-\tau_a, \mathbf{x}_a)) \right. \\
    &\quad - \frac{3}{4}(E^2_a(t, \mathbf{x}=0) - E^2_a(t-\tau_a, \mathbf{x}_a)) \\
    &\quad \left. + \frac{5}{8}(E^3_a(t, \mathbf{x}=0) - E^3_a(t-\tau_a, \mathbf{x}_a)) + \dots \right] ,
\end{align}
where $\tau_a = t_\text{obs} - t_\text{em}$ is the light travel
between the Earth and the pulsar $a$. In \autoref{eq:za}, the
linear order is responsible only for the power spectrum, while for
the trispectrum the second and third orders are also responsible.

The power spectrum $P$ of the SGWB is usually
\begin{equation} \label{eq:power}
    \langle h_{A_1}(f_1, \hat{\mathbf{n}}_1) h_{A_2}(f_2, \hat{\mathbf{n}}_2) \rangle = \delta(f_1 + f_2) \delta^{(2)}(\hat{\mathbf{n}}_1 - \hat{\mathbf{n}}_2) \delta_{AA'} P(f_1) .
\end{equation}
The first two delta functions are due to the requirement of time
and space translationally invariant. Similarly, there should be
$\delta(f_1+f_2+f_3+f_4)\delta(f_1\hat{\mathbf{n}}_1 +
f_2\hat{\mathbf{n}}_2 + f_3\hat{\mathbf{n}}_3 +
f_4\hat{\mathbf{n}}_4)$ present in the stationary 4-point correlation
function $\langle h_{A_1}(f_1, \hat{\mathbf{n}}_1) h_{A_2}(f_2,
\hat{\mathbf{n}}_2) h_{A_3}(f_3, \hat{\mathbf{n}}_3) h_{A_4}(f_4,
\hat{\mathbf{n}}_4) \rangle$.
% { \color{red}
This will lead to a folded (a.k.a. flatten) shape in the Fourier space, i.e. all four $\vec{k} = f\hat{\mathbf{n}}$ wave vectors are aligned and connected head to tail, as shown in \autoref{fig:folded}.
% }
\begin{figure}[htb]
    \centering
\begin{tikzpicture}[x=0.75pt,y=0.75pt,yscale=-1,xscale=1]
\draw    (209,48) -- (22,48) ;
\draw [shift={(20,48)}, rotate = 360] [color={rgb, 255:red, 0; green, 0; blue, 0 }  ][line width=0.75]    (10.93,-3.29) .. controls (6.95,-1.4) and (3.31,-0.3) .. (0,0) .. controls (3.31,0.3) and (6.95,1.4) .. (10.93,3.29)   ;
\draw    (20,48) -- (87,52.58) ;
\draw [shift={(89,52.72)}, rotate = 183.91] [color={rgb, 255:red, 0; green, 0; blue, 0 }  ][line width=0.75]    (10.93,-3.29) .. controls (6.95,-1.4) and (3.31,-0.3) .. (0,0) .. controls (3.31,0.3) and (6.95,1.4) .. (10.93,3.29)   ;
\draw    (89,52.72) -- (146,52.72) ;
\draw [shift={(148,52.72)}, rotate = 180] [color={rgb, 255:red, 0; green, 0; blue, 0 }  ][line width=0.75]    (10.93,-3.29) .. controls (6.95,-1.4) and (3.31,-0.3) .. (0,0) .. controls (3.31,0.3) and (6.95,1.4) .. (10.93,3.29)   ;
\draw    (148,52.72) -- (207.01,48.15) ;
\draw [shift={(209,48)}, rotate = 175.58] [color={rgb, 255:red, 0; green, 0; blue, 0 }  ][line width=0.75]    (10.93,-3.29) .. controls (6.95,-1.4) and (3.31,-0.3) .. (0,0) .. controls (3.31,0.3) and (6.95,1.4) .. (10.93,3.29)   ;

\draw (102,28.4) node [anchor=north west][inner sep=0.75pt]    {$\vec{k}_{4}$};
% Text Node
\draw (44,56.4) node [anchor=north west][inner sep=0.75pt]    {$\vec{k}_{1}$};
% Text Node
\draw (101,57.07) node [anchor=north west][inner sep=0.75pt]    {$\vec{k}_{2}$};
% Text Node
\draw (165,59.07) node [anchor=north west][inner sep=0.75pt]    {$\vec{k}_{3}$};
\end{tikzpicture}
    \caption{Schematic illustration of the wavevector $\vec{k}_i = f_i\hat{\mathbf{n}}_i$ of the gravitational waves that follow the folded shape focused in this paper. We also consider the permutation between indexes $i$.}
    \label{fig:folded}
\end{figure}
Furthermore, we consider the SGWB with an unpolarized trispectrum
for simplicity:
\begin{equation} \label{eq:tri}
\langle h_{A_1}(f_1, \hat{\mathbf{n}}_1) h_{A_2}(f_2,
\hat{\mathbf{n}}_2) h_{A_3}(f_3, \hat{\mathbf{n}}_3) h_{A_4}(f_4,
\hat{\mathbf{n}}_4) \rangle = \alpha \left( \prod_{i=1}^3
\delta^{(2)}(\hat{\mathbf{n}}_i - \hat{\mathbf{n}}_4) \delta(f_4 +
3f_i) \delta_{A_i A_4} \times P(f_4) + \text{perms.} \right) ,
\end{equation}
which has one parameter $\alpha$ characterizing the amplitude of the
trispectrum.
% The second is the one studied in Ref.\cite{Tasinato:2022xyq}:
% \footnote{We have added a permutation here compared to Ref.\cite{Tasinato:2022xyq}, thus the definition of the parameters $\kappa_1, \kappa_2$ is slightly different.}
% \begin{multline} \label{eq:tri2}
%     \langle h_{A_1}(f_1, \hat{\mathbf{n}}_1) h_{A_2}(f_2,
%     \hat{\mathbf{n}}_2) h_{A_3}(f_3, \hat{\mathbf{n}}_3) h_{A_4}(f_4,
%     \hat{\mathbf{n}}_4) \rangle = \\
%     \left( \prod_{i=1}^3
%     \delta^{(2)}(\hat{\mathbf{n}}_i - \hat{\mathbf{n}}_4) \delta(f_4 +
%     3f_i) \left(\kappa_1 \delta_{A_1A_2}\delta_{A_3A_4} + \kappa_2(1-\delta_{A_1A_2})(1-\delta_{A_3A_4})\delta_{A_1A_3}\delta_{A_2A_4}\right) \times P(f_4) + \text{perms.} \right) ,
% \end{multline}
We assume the parameters $\alpha$
% , $\kappa_1$ and $\kappa_2$
here are frequency-independent for simplicity. 
It should be
mentioned that the trispectrum may also impact
the cosmic variance of the observation
\cite{Allen:2022dzg,Bernardo:2022xzl,Bernardo:2023bqx,Bernardo:2023pwt,Bernardo:2023zna},
as it is a four-point function of the GW amplitudes.
% { \color{red}
Although the cosmic variance may affect detectability for non-Gaussianity, we expect it to drop with observation time for broadband signals as the frequency of the signal becomes more distinct.
Here, we follow NANOGrav's analysis and ignore its effects.
% }

\begin{figure}
    % \begin{subfigure}{0.49\textwidth}
    \includegraphics{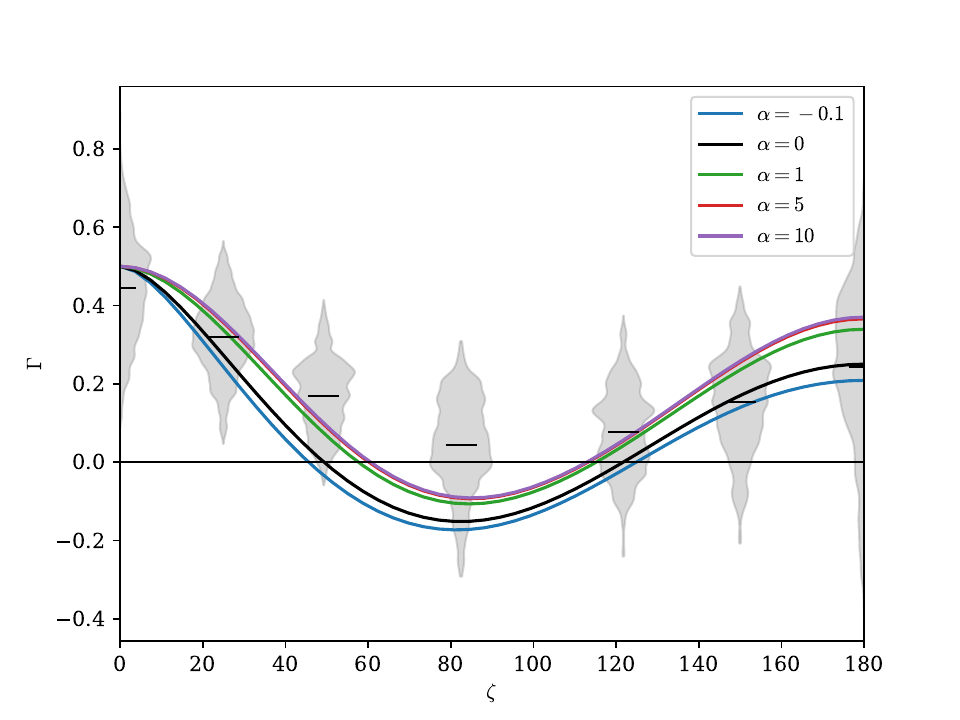}
    % \end{subfigure}
    % \begin{subfigure}{0.49\textwidth}
    % \includegraphics[width=\linewidth]{ORF2.pdf}
    % \end{subfigure}
\caption{The dependence of the normalized ORF on the relative
angle $\zeta$ between pulsars for our parameterization. We also
show Bayesian reconstruction with the cubic spline ORF based on
the NANOGrav 15-year data \cite{NANOGrav:2023gor} as the gray
violins plot.}
    \label{fig:ORF}
\end{figure}

The 2-point correlator $\langle z_a(t) z_b(t) \rangle$ between two
pulsar $a,b$ corresponds to \cite{Tasinato:2022xyq}
\begin{equation} \label{eq:2pt_correlator}
    \langle z_a(t) z_b(t) \rangle = \beta \ \Gamma(\zeta_{ab}) \int \dd f P(f) ,
\end{equation}
where $\beta$ is a normalization factor and $\zeta_{ab}$ is the
relative angle between the two pulsars. The ORF can be obtained
using \autoref{eq:za} and \autoref{eq:power}
% for the first parameterization:
and \autoref{eq:tri}:
\begin{equation} \label{eq:ORF}
\Gamma(\zeta_{ab}) \propto \sum_{A=+,\times} \int_{S^2} \dd
\hat{\mathbf{n}} \frac{\bar{E}_a^{A} \bar{E}_b^{A} + 4\alpha
\left(
\frac{9}{16}\bar{E}_a^{A}\bar{E}_a^{A}\bar{E}_b^{A}\bar{E}_b^{A} +
\frac{5}{8}(\bar{E}_a^{A}\bar{E}_a^{A}\bar{E}_a^{A}\bar{E}_b^{A} +
\bar{E}_a^{A}\bar{E}_b^{A}\bar{E}_b^{A}\bar{E}_b^{A}) \right)}{(1
+ \hat{\mathbf{n}}_a \cdot \hat{\mathbf{n}} )( 1 +
\hat{\mathbf{n}}_b \cdot \hat{\mathbf{n}})} ,
\end{equation}
% while for the second parameterization:
% \begin{multline} \label{eq:ORF2}
%     \Gamma(\zeta_{ab}) \propto \sum_{A_i=+,\times} \int_{S^2} \dd \hat{\mathbf{n}} \frac{1}{(1 + \hat{\mathbf{n}}_a \cdot \hat{\mathbf{n}} )( 1 + \hat{\mathbf{n}}_b \cdot \hat{\mathbf{n}})} (\delta_{A_1 A_2} \bar{E}_a^{A_1} \bar{E}_b^{A_2} +  4 \left(\kappa_1 \delta_{A_1A_2}\delta_{A_3A_4} +\right. \\
%     \left. \kappa_2(1-\delta_{A_1A_2})(1-\delta_{A_3A_4})\delta_{A_1A_3}\delta_{A_2A_4}) \left( \frac{9}{16} \bar{E}_a^{A_1}\bar{E}_a^{A_2}\bar{E}_b^{A_3}\bar{E}_b^{A_4} + \frac{5}{8} (\bar{E}_a^{A_1}\bar{E}_a^{A_2}\bar{E}_a^{A_3}\bar{E}_b^{A_4} + (a \leftrightarrow b)) \right) \right),
% \end{multline}
where $\bar{E}_a^{A} = \mathbf{e}_{ij}^A \hat{\mathbf{n}}_a^i
\hat{\mathbf{n}}_a^j$, $\mathbf{e}_{ij}^A$ is the basis for the
two kinds of polarization tensors. The contribution corresponding
to the bispectrum vanishes after integration, thus the leading
order except the linear order is the contribution from the
trispectrum. This can be integrated by specifying the angle
parameter. We show the result for some parameters in
\autoref{fig:ORF}. When $\alpha \rightarrow 0$, only the first
term in the numerator survives and we restore the original
Hellings-Downs curve \cite{Hellings:1983fr}.
Moreover,
by comparing the case with $\alpha=5$ and $\alpha=10$,
we can find that when $\alpha$ is sufficiently large,
the second term on the numerator
of \autoref{eq:ORF} dominates and the ORF asymptotically develops
another shape.
% The behavior of $\kappa_1$ is similar as it also contains the contribution
% of unpolarized combination.
% Comparing the cases of $\kappa_2=0$ and $\kappa_2=1$ with $\kappa_1=0$, we can see that a positive $\kappa_2$ does not make a significant impact on the ORF.
% This is also true for the case where $\kappa_1$ is positive, and we can see that the cases $\kappa_1=1, \kappa_2=1$ have the similar asymptotic shape as the large positive $\kappa_1$.
% However, when $\kappa_2$ is negative, it leads to a dramatically different ORF shape.
%We numerically pre-calculate the ORF and interpolate it during the fitting.

\subsection{NANOGrav 15-yr data set}

Following the approach of NANOGrav \cite{NANOGrav:2023icp}, we
only make use of 67 pulsars with observational timespans over 3
years. The pulse times of arrival (TOAs) consist of 4 parts:
\begin{equation}
    \mathbf{t} = \mathbf{t}_\text{det} + \mathbf{t}_\text{WN} + \mathbf{t}_\text{RN} + \mathbf{t}_\text{SGWB}
\end{equation}
where $\mathbf{t}_\text{det}$ is the deterministic part,
$\mathbf{t}_\text{WN}$ is the white noise due to measurement
uncertainties, $\mathbf{t}_\text{RN}$ is the intrinsic red noise
for each pulsars and $\mathbf{t}_\text{SGWB}$ is the common red
noise due to the SGWB. The covariance matrix of the common red
noise between pulsars is described by the ORF (\autoref{eq:ORF}),
while the dependence of the intrinsic red noise on the frequencies
is modeled as a power-law spectrum with an amplitude $A_\text{RN}$
and the spectral index $\gamma_\text{RN}$. The common red noise is
also modeled as a power-law spectrum with an amplitude
$A_\text{GW}$ and the spectral index $\gamma_\text{GW}$. The
reference frequency is 1 yr$^{-1}$. They are expressed as sine and
cosine components with frequencies $f_i = i/T$, where $T$ is the
observational timespan of the entire data set. We choose the first
14 modes for the common red noise and the first 30 modes for the
intrinsic red noise, following NANOGrav
\cite{NANOGrav:2020bcs,NANOGrav:2023icp}.

%Modes with $f = \frac{3}{T}, \frac{6}{T}, \frac{12}{T}$ of the
%common red noise are not included in the likelihood as they may
%have non-trivial covariance with other modes in our case.
%\footnote{See Ref.\cite{Lentati:2014hja} for an attempt to
%construct a likelihood of non-Gaussian noise.}

\begin{table}[]
    \centering
    \footnotesize
    \begin{tabular}{lp{7cm}lp{6cm}} \hline \hline
        parameter & description & prior & comments \\ \hline
        \multicolumn{4}{c}{intrinsic red noise} \\
        $A_\text{RN}$ & red-noise power-law amplitude & log-Uniform [-20, -11] & one parameter per pulsar\\
        $\gamma_\text{RN}$ & red-noise power-law spectral index & Uniform [0, 7] & one parameter per pulsar \\ \hline
        \multicolumn{4}{c}{common processes} \\
        $A_\text{GW}$ & SGWB strain  power-law amplitude & log-Uniform [-18, -11] & one parameter for PTA \\
        $\gamma_\text{GW}$ & SGWB power-law spectral index & Uniform [0, 7] & one parameter for PTA,\newline fixed to $\frac{13}{3}$ for signals from SMBHBs \\
        $\alpha$ & unpolarized SGWB trispectrum amplitude\newline (\autoref{eq:tri}) & Uniform [-10, 10] & one parameter for PTA \\
        % $\kappa_1$ & SGWB trispectrum parameter (\autoref{eq:tri2}) & Uniform [-5, 5] & one parameter for PTA \\
        % $\kappa_2$ & SGWB trispectrum parameter (\autoref{eq:tri2}) & Uniform [-5, 5] & one parameter for PTA \\
        \hline \hline
    \end{tabular}
    \caption{Prior distributions used in our analysis.}
    \label{tab:prior}
\end{table}

To explore the parameter space, we perform a Bayesian analysis
with Markov chain Monte Carlo (MCMC) sampling. The white noise
parameters (EFAC, EQUAD, and ECORR) are analyzed initially for
individual pulsars. Afterwards, they are fixed and the common red
noise parameters are varied with the intrinsic red noise
parameters. The prior employed for these parameters is summarised
in \autoref{tab:prior}.
We consider the two cases of free $\gamma_\text{GW}$ and $\gamma_\text{GW}=13/3$, i.e., signals from SMBHBs.
In order to compare with the model without
non-Gaussianity, we also compute the Bayes factor between two
models:
\begin{equation}
    \mathcal{B}_\text{A,B} = \frac{\mathcal{Z}_A}{\mathcal{Z}_B} ,
\end{equation}
where $\mathcal{Z}_A$ is the Bayes evidence for model A.
We utilize the \texttt{enterprise, enterprise\_extensions} packages \cite{enterprise}
and the \texttt{PTMCMCSampler} package \cite{justin_ellis_2017_1037579} to perform the MCMC sampling.
Following NANOGrav, We perform the product-space sampling~\cite{carlin1995bayesian,godsill2001relationship} to estimate the Bayes factor and use statistical bootstrapping~\cite{efron1986bootstrap} to estimate its uncertainties.
% The posterior distributions are plotted with the \texttt{GetDist} package \cite{Lewis:2019xzd}.

% @misc{enterprise,
%   author       = {Justin A. Ellis and Michele Vallisneri and Stephen R. Taylor and Paul T. Baker},
%   title        = {ENTERPRISE: Enhanced Numerical Toolbox Enabling a Robust PulsaR Inference SuitE},
%   month        = sep,
%   year         = 2020,
%   howpublished = {Zenodo},
%   doi          = {10.5281/zenodo.4059815},
%   url          = {https://doi.org/10.5281/zenodo.4059815}
% }
% @misc{justin_ellis_2017_1037579,
%   author       = {Justin Ellis and
%                   Rutger van Haasteren},
%   title        = {jellis18/PTMCMCSampler: Official Release},
%   month        = oct,
%   year         = 2017,
%   doi          = {10.5281/zenodo.1037579},
%   url          = {https://doi.org/10.5281/zenodo.1037579}
% }

% \section{Results and Discussion} \label{sec:result}

\begin{figure}
     \begin{subfigure}{0.49\textwidth}
         \centering
         \includegraphics[width=\textwidth]{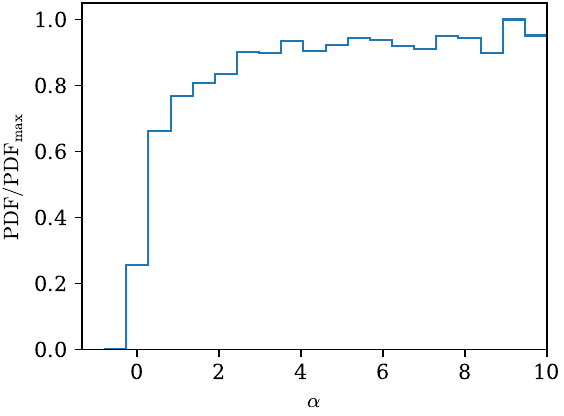}
         \caption{Free $\gamma_\text{GW}$.}
     \end{subfigure}
     \begin{subfigure}{0.49\textwidth}
         \centering
         \includegraphics[width=\textwidth]{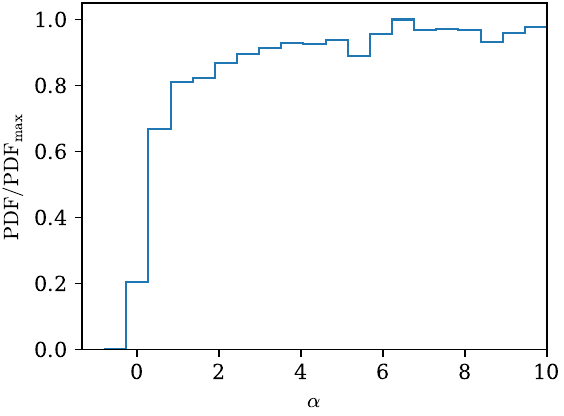}
         \caption{$\gamma_\text{GW}$ fixed to $\frac{13}{3}$ for signals from SMBHBs.}
     \end{subfigure}
    \caption{Binned posterior distributions of the parameter $\alpha$ under two $\gamma_\text{GW}$ choices.}
    \label{fig:alpha}
\end{figure}

\section{Results and Discussion} By allowing $\alpha$ to vary, we
find the median and 90\% credible region of the GWB amplitude
$A_\text{GW}=5.2^{+4.5}_{-2.7}\times 10^{-15}$ and the spectrum
index $\gamma_\text{GW} = 3.3 \pm 0.7$,
% and by allowing $\kappa_1, \kappa_2$ to vary,
% we find $A_\text{GW}=4.1^{+3.7}_{-1.6}\times10^{-15}$
% and $\gamma_\text{GW}=3.6^{+0.5}_{-0.8}$
which are consistent with the NANOGrav results
\cite{NANOGrav:2023gor}.
Besides, when $\gamma_\text{GW}$ is fixed to $13/3$, we find $A_\text{GW} = 2.0^{+0.7}_{-0.5}\times 10^{-15}$.

We show the binned posterior distribution of $\alpha$ in
\autoref{fig:alpha}. The first feature we can find is that
negative $\alpha$ is strongly disfavored. This is because a
negative $\alpha$ would correspond to a shape that deviates
dramatically from the HD curve. In fact, the right-hand side of
\autoref{eq:ORF} becomes negative when $\alpha$ is sufficiently
small, which prevents \autoref{eq:2pt_correlator} from being
normalized. Another significant feature is that when $\alpha$ is
sufficiently large ($\alpha \gtrsim$ 4), the posterior becomes
flat. This is a reflection of the asymptotic behavior of the ORF
\autoref{eq:ORF} mentioned above.
In the case of a free $\gamma_\text{GW}$,
the Bayes factor between the
model with a folded non-Gaussianity parameterized by $\alpha$ and
the model without non-Gaussianity in the case is
\begin{equation}
\mathcal{B}_\text{unporlarized trispectrum,HD} = 1.68 \pm 0.01 \, .
\end{equation}
% { \color{red}
Although the Bayes factor is greater than 1, located on the side of the model with non-Gaussianity, there is no evidence for non-Gaussianity according to the Jeffreys scale~\cite{Jeffreys:1939xee}.
% }
% which suggests the model with non-Gaussianity is slightly favored according to the Jeffreys scale \cite{Jeffreys:1939xee}.
And while $\gamma_\text{GW}$ is fixed at $13/3$, we find that the Bayes factor is slightly raised:
\begin{equation}
    \mathcal{B}_\text{unporlarized trispectrum,HD} = 1.78 \pm 0.01.
\end{equation}
These
results can also be confirmed by the comparison between the ORF
calculated with \autoref{eq:ORF} and the Bayesian reconstructed
ORF shown in \autoref{fig:ORF}, where we can find that the middle
three bins prefer a smaller $\Gamma$ with respect to the HD curve,
and a positive folded non-Gaussianity can compensate for this
difference\footnote{
% \color{red}
Such deviations may also appear in other models or be due to some factors, e.g. unmodelled noise with monopolar or dipolar correlations, the cosmic variance (e.g.~\cite{Allen:2022dzg,Bernardo:2022xzl,Bernardo:2023bqx,Bernardo:2023pwt,Bernardo:2023zna}) or modified gravity (e.g.~\cite{Bi:2023ewq,Wu:2023rib}).
}.

% The posterior distribution of $\kappa_1, \kappa_2$ is shown in \autoref{fig:alpha}.
% The behavior of $\kappa_1$ is similar to $\alpha$,
% due to their similar effects on ORF.
% Negative $\kappa_2$ is disfavored as it will lead to an ORF shape that deviates dramatically from the observational constraints.
% Increasing $\kappa_2$ will only change the ORF slightly, so the posterior has no significant dependence on $\kappa_2$ when it is positive.
% The case without non-Gaussianity ($\kappa_1=0, \kappa_2=0$) is located in the $2\sigma$ range.

In certain sense, we are testing the ability of PTA experiments in
searching for the non-Gaussianity of SGWB through the non-linear
effect of the non-Gaussianity on the ORF. In the light of our
result, our approach and its further extension might bring the
unforeseen insights into the origin of the corresponding SGWB.

%Finally, it should be noted that we are testing the ability of PTA
%experiments to constrain non-Gaussianity with real data, are not
%considering here an estimator that utilizes all the information in
%the PTA experiments, since we only the impact of the
%non-Gaussianity on the overlap reduction function. There are other
%pieces of information such as the non-trivial covariance between
%time residuals and higher points correlators.
%Meanwhile, we only consider the unpolarized trispectrum, and other
%polarization modes can bring different effects. We leave them to
%future studies.

\begin{acknowledgments}
JQJ thanks Hai-Long Huang, Ao Guo, and Yan-Chen Bi for useful
discussions. YSP is supported by NSFC, No.12075246, National Key
Research and Development Program of China, No. 2021YFC2203004, and
the Fundamental Research Funds for the Central Universities.
We acknowledge the use of high performance computing services provided by the International Centre for Theoretical Physics Asia-Pacific cluster.
\end{acknowledgments}

\bibliography{refs}

\end{document}